\documentclass[10pt, conference]{IEEEtran}
\usepackage{braket}
\usepackage{color, soul,xcolor}
\usepackage{float}
\usepackage{amsmath}

\usepackage{graphicx}
\usepackage{subfig}
\usepackage{booktabs}
\usepackage[hidelinks]{hyperref}

\usepackage{cite}

\newcommand\copyrighttext{%
  \footnotesize \textcopyright 2021 IEEE.  Personal use of this material is permitted.  Permission from IEEE must be obtained for all other uses, in any current or future media, including reprinting/republishing this material for advertising or promotional purposes, creating new collective works, for resale or redistribution to servers or lists, or reuse of any copyrighted component of this work in other works. DOI: \href{https://doi.org/10.1109/NANOARCH53687.2021.9642249}{10.1109/NANOARCH53687.2021.9642249}}

\begin{document}
\title{Error Analysis of the Variational Quantum Eigensolver Algorithm}

\author{
\thanks{\copyrighttext . This work was partially funded by the Carl Zeiss foundation.}

\begin{tabular}{@{}c@{\qquad\quad}c@{}}
\multicolumn{2}{c}{Sebastian Brandhofer$^{1}$ \qquad Simon Devitt$^2$ \qquad Ilia Polian$^{1}$}\\[0.5cm]
    \begin{tabular}{@{}c@{}}
\normalsize
    $^1$Institute of Computer Architecture and Computer Engineering,\\
\normalsize
    Center for Integrated Quantum Science and Technology (IQ$^\text{ST}$)\\
\normalsize
    University of Stuttgart, Stuttgart, Germany \\
\normalsize
    \{sebastian.brandhofer\;$|$\;ilia.polian\}@informatik.uni-stuttgart.de
    \end{tabular}
&
    \begin{tabular}{@{}c@{}}
\normalsize
    $^2$Centre for Quantum Software and Information \\
\normalsize
    University of Technology Sydney, Sydney, Australia \\
\normalsize
    simon.devitt@uts.edu.au\hspace{0cm}\\
    \\
    \end{tabular}

\end{tabular}
\vspace{-0.5ex}
}

\IEEEoverridecommandlockouts

\IEEEpubidadjcol
\maketitle

\begin{abstract}
Variational quantum algorithms have been one of the most intensively studied applications for near-term quantum computing applications.
The noisy intermediate-scale quantum (NISQ) regime, where  small enough algorithms can be run successfully on noisy quantum computers expected during the next 5 years, is driving both a large amount of research work and a significant amount of private sector funding. 
Therefore, it is important to understand whether variational algorithms are effective at successfully converging to the correct answer in presence of noise.
We perform a comprehensive study of the variational quantum eigensolver (VQE) and its individual quantum subroutines. 
Building on asymptotic bounds, we show through explicit simulation that the VQE algorithm effectively collapses already when single errors occur during a quantum processing call.
We discuss the significant implications of this result in the context of being able to run any variational type algorithm without resource expensive error correction protocols.
\end{abstract}

\IEEEpeerreviewmaketitle

\section{Introduction}

Variational quantum algorithms are considered by many as the first candidates to solve problems in chemistry and material sciences that are classically intractable and practically relevant on near term quantum computers.
Even though today's quantum computers are affected by high levels of noise, variational algorithms are hoped to produce useful results due to their hybrid structure, where short quantum computations (ansatzes) are intertwined with error-free classical optimization steps.
This optimism is reinforced by experimental demonstrations of the variational quantum eigensolver (VQE) algorithm for small-scale computational chemistry problems~\cite{34, 10, 2}~on molecules such as H$_2$ or LiH.
These examples can easily be calculated on a classical computer.
If VQE can be extended to larger molecules, we can expect revolutions in domains ranging from drug design to development of fertilizers~\cite{34, 20, 23}. 

This paper studies the question whether VQE will scale to larger problem instances assuming the error rates of today’s and tomorrow’s \emph{noisy intermediate-scale quantum} (NISQ) computers. 
NISQ devices are currently associated with an error rate of order 1\%, and their limited size of less than 100 qubits prohibit active error correction protocols~\cite{1}.
We show that already for medium-sized problem instances, one error in 100 VQE quantum computations is sufficient to ``derail'' the algorithm and break its convergence; modern quantum computers will produce much higher error rates.
This is in stark contrast to small VQE instances where the individual quantum computations are smaller and have a fair chance to run without significant errors on today's hardware.
For larger problem instances in chemistry and optimization, however, often larger VQE quantum computations are required that enable the traversal of a larger state space~\cite{13, 17, 15, 16, 14}.

\begin{figure}[b]
\vspace{-3ex}
  \centering
  \includegraphics[width=0.99\linewidth]{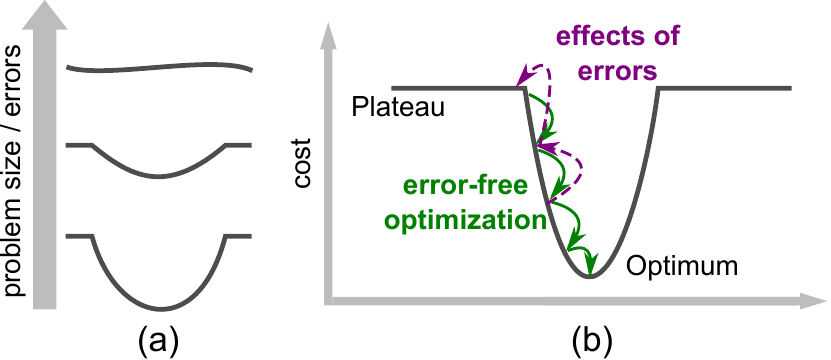}

  \caption{a. Noise-induced barren plateau (adapted from \cite{32}); b. Convergence behavior of VQE subject to errors \label{11}}
\end{figure}

Our results are in line with a recent theoretical finding that variational algorithms, including VQE, are prone to \emph{noise-induced barren plateaus}~\cite{32}.
The gradients calculated in the noise-affected part are deteriorating, transforming an optimization landscape with a minimum into a flat landscape where the minimum cannot be reached.
This effect increases with problem size and error strength (fig.~\ref{11}a).
Fig.~\ref{11}b summarizes our view of the problem: if the minimum is located in a small part of the high-dimensional parameter space, a large number of steps (indicated by green arrows) are needed to descend into it.
An error during one of these steps may increase the distance to the minimum, potentially bringing the algorithm to a plateau from which it can not converge to the minimum.

This work extends the insight into the error behavior of VQE through numerical simulations subject to errors by:
\begin{itemize}
    \item Giving a closer bound of a typical VQE application on acceptable errors and quantum computation sizes compared to asymptotic bounds in~\cite{32}.

    \item Showing previously unexplored requirements of VQE on errors in quantum computers.
    \item Exposing the emergence of noise-induced barren plateaus with an increasing error rate.
    \item Relating the impact of Pauli errors on the overall VQE solution and the error characteristics of the quantum computations performed by VQE on VQE convergence.
\end{itemize}
The remainder of the work is organized as follows.
The background on quantum computing and the variational quantum eigensolver is provided in section~\ref{1}~and~\ref{13}~respectively.
Section~\ref{4}~discusses related work and in section~\ref{5}~the experimental setup is described.
We present the results of our simulations in section~\ref{7}~and outline the implications on computing variational algorithms on NISQ computers in section~\ref{6}.

\section{Quantum Computing}\label{1}
Quantum computing promises an exponential speed up for classically intractable problems in e.g. cryptography~\cite{33}~or chemistry~\cite{11, 34}~by manipulating and measuring quantum states.
A quantum algorithm specifies on an abstract level how such a quantum state must be manipulated and measured to yield the solution to a computational problem such as integer factorization.
From a quantum algorithm, a quantum circuit is derived that specifies the operations (and their order) a quantum computer must perform to follow the steps described in the quantum algorithm.
These operations are called quantum gates.
A quantum computer can store, manipulate and measure quantum states depending on external control signals that are specified by a quantum circuit. 

Noisy quantum computations, prevalent in NISQ devices, corrupt the quantum state manipulated by a quantum computer, effectively shortening reasonable computation times and leading to wrong solutions.
Noise can be represented as errors on the logical level, e.g. Pauli errors~\cite{1}, or modeled based on adverse interactions on the physical level.
Simulating the impact of noise on quantum circuits is typically done using the density operator formalism that allows to accurately represent noise without the need for Monte Carlo sampling but requires exponentially more space than a state vector simulator~\cite{27}.
State vector simulators can be used to simulate noise on the logical level, i.e. Pauli errors~\cite{5}, and can handle twice as large quantum circuits at the same memory footprint compared to density operator simulators~\cite{27}, but require statistical sampling over possible error locations. 
Pauli errors were shown to accurately describe the impact of noise on quantum computations performed by Google's Sycamore quantum computer~\cite{9} and consequently represent the {\em best case} error scenario for well engineered chipsets, suitable for quantum computation.

\section{Variational Quantum Eigensolver}\label{13}
Using the Variational Quantum Eigensolver (VQE) algorithm, the minimum eigenvalue $\lambda_{min}$ of a target hermitian operator $H$ can be computed.
This yields promising applications on large operators for chemistry simulations, material sciences, drug design and combinatorial optimization that are intractable to solve with classical computers~\cite{34, 20, 16, 23}.
VQE is able to target large operators, while only executing relatively short quantum circuits.
Therefore, there is hope that VQE can yield results of industry interest on NISQ computers.

\begin{figure}[ht!]
  \centering
  
  \includegraphics[width=.75\linewidth]{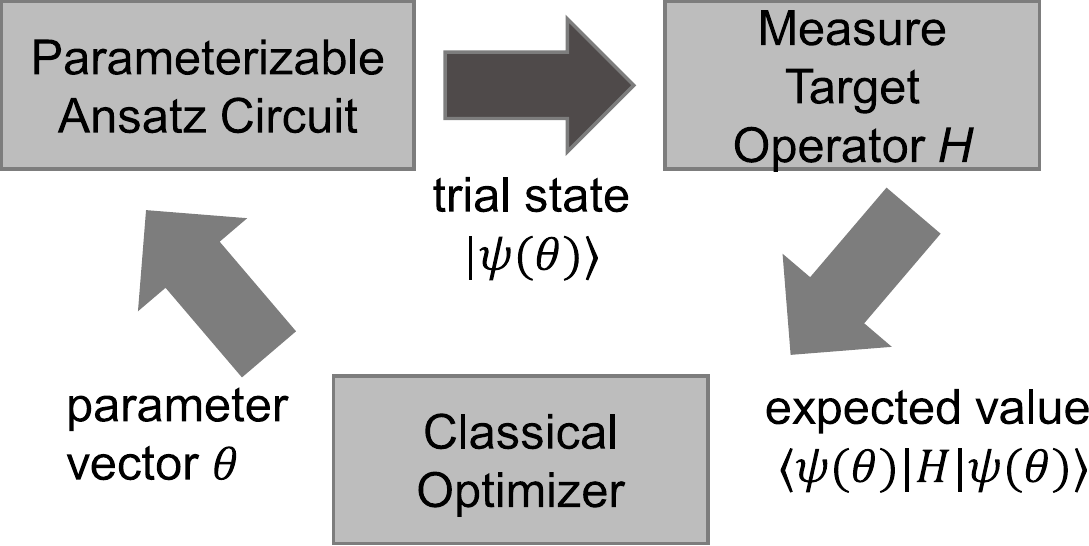}
  \caption{Steps of the Variational Quantum Eigensolver \label{14}}
\end{figure}

VQE iterates over the individual computation steps shown in fig.~\ref{14}~until the computation converges to $\lambda \geq \lambda_{\min}$.
First, a parameterizable circuit computes a trial state $\ket{\psi(\theta)}$.
This trial state is then measured with respect to $H$.
The measurement result $\lambda_{\theta} =\bra{\psi(\theta)}H\ket{\psi(\theta)}$ is a real value and is an input to the classical optimizer.
Finally, the classical optimizer chooses a new parameter vector based on $\lambda_{\theta}$.
The initial parameter vector can be randomly chosen or motivated by the problem domain, i.e. set such that the initial trial state is expected to be close to the optimum\cite{6}.
The optimizer terminates the computation if a set convergence rate, the maximum runtime or the maximum number of trial states is reached.
One VQE iteration consists of the preparation of a trial state, the measurement of a target operator and the selection of the next parameter.
\subsection{Ansatz Circuits}
\begin{figure}[tb!]
  \centering
  
  \includegraphics[width=\linewidth]{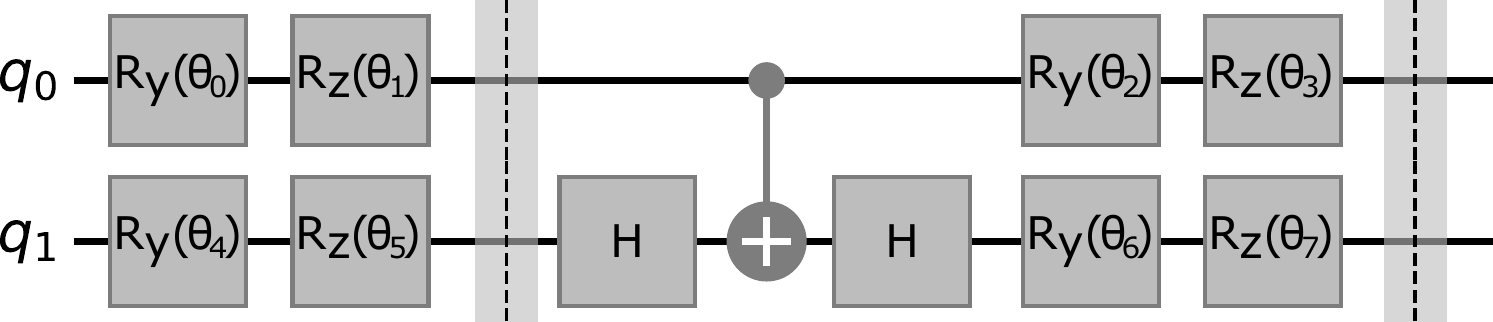}
  \caption{A 2-qubit RYRZ ansatz circuit with depth 1 and 8 parameters. The gates bounded by dashed lines are repeated for circuits with depth larger than one.\label{9}}
  \vspace{-2ex}
\end{figure}

An ansatz circuit computes a state based on a vector of provided parameters.
For example, the 2-qubit RYRZ ansatz circuit shown in figure~\ref{9} has depth 1 and 8 parameters that define single-qubit rotations.
Ansatz circuits must be able to approximate a trial state that corresponds to the minimal eigenvalue with the finite number of parameters supported by classical optimizers.
In addition, ansatz circuits can be designed to be efficiently computed by a specific NISQ computer~\cite{10} or to efficiently represent certain problems, e.g. in chemistry~\cite{29}.
The structure of the utilized ansatz circuit can also be further adapted during the execution of VQE~\cite{24}.

Errors in the ansatz circuit computation can change the output state of the circuit and may corrupt target operator measurements.
This can lead to an erroneous mapping of parameters to measurement results, which has an impact on the optimization trajectory; the optimizer may erroneously chose a subsequent vector of parameters that increases the distance to the optimum.

\subsection{Target Operator Measurement}\label{13_operator}
Depending on the underlying technology, an arbitrary hermitian operator $H$ can often not be measured by a quantum computer directly\cite{9}.
Instead the target operator is decomposed into a weighted sum of Pauli operators that are measured individually, i.e. one adapted ansatz circuit (measurement circuit) must be computed and measured for every set of decomposed Pauli operators.
The decomposition of the target operator must yield a polynomial number of Pauli operator measurements for efficient eigenvalue computation~\cite{34}.
Hermitian operators whose minimum eigenvalue corresponds to the ground state energy, for instance, incur $O(n^{4})$ Pauli operator measurements~\cite{0}, where $n$ is the number of qubits in the ansatz circuit.
The number of required Pauli operator measurements can be reduced by exploiting the commutation relations of the individual Pauli operators~\cite{22, 7}.

The Pauli operator decomposition and corresponding measurements add a significant overhead to the computation of VQE that may prevent the practicality of VQE for large-scale target operators~\cite{22, 4}.
This overhead also leads to a larger total number of errors that must be handled by VQE as more computations must be performed on the quantum computer.
Errors during the measurement of a target operator have a similar impact on VQE as errors in the ansatz circuit; the classical optimizer is confronted with an erroneous mapping of parameters to measurement result that may lead to incorrect optimization directions.

\subsection{Classical Optimizer}
The classical optimizer sets the next parameter vector based on the current parameters and a chosen optimization step:
\begin{equation}
    \theta_{t+1} = \theta_{t} + \alpha_{t} d_{t}
\end{equation}
where $d_{t}$ is an optimization direction in the parameter space and $\alpha_{t}$ is the optimization step size.
Numerous strategies exist for exploring the parameter space through optimization steps.
Depending on the available information, $d_t$ can be set by evaluating the gradient~\cite{19}, a gradient estimation~\cite{31,30} or the history of data points~\cite{8, 36}~at $\theta_t$.
The step size $\alpha_t$ can be a fixed value or adapted during the optimization based on the expected improvement or a predefined schedule~\cite{30}.

\section{Related Work}\label{4}
In the recent past, computations of variational quantum algorithms subject to errors were investigated through theoretical analyses~\cite{32},
simulations~\cite{28, 25, 26, 21}~and experiments~\cite{34, 2, 10}~on quantum computers.

VQE was experimentally demonstrated to find optimal or near-optimal solutions to small-scale chemistry applications that require ansatz circuits with up to 12 qubits and a small number of quantum gates per qubit~\cite{2, 10, 34}.
For larger ansatz circuits, theoretical analyses have shown the existence of noise-induced barren plateaus~\cite{32}.
These analyses provide general statements about specific types of 
ansatz circuits and variational quantum algorithms without considering 
the overhead of the target operator measurements.
However, this generality also leads to asymptotic bounds that are tightened through explicit numerical simulations and by explicitly considering target operator measurements in this work on a typical setup of the variational quantum eigensolver for chemistry applications.

Numerical simulations in the related work mostly focus on the impact of errors on the overall solution obtainable by VQE for combinatorial~\cite{28, 25}~or chemistry problems~\cite{10, 21, 26}.
In contrast, our simulations offer an insight into the emergence of noise-induced barren plateaus and their relation to the error characteristics of the employed ansatz circuit.
We consider both: impact of errors on the individual ansatz circuits and convergence behavior as well as the solution of error-affected VQE.
Furthermore, the impact of errors is additionally evaluated as a function of the number of erroneous ansatz circuits by simulating a fixed number of ansatz circuit computations subject to one random Pauli error.
Knowing the acceptable ratio of erroneous ansatz circuits exposes previously unexplored requirements on quantum computers that are essential to estimate success of a VQE run.
Moreover this is a generic and technology-independent value that does not only apply to a specific technology investigated in~\cite{28, 10}.

\section{Methodology}\label{5}
For our numerical simulations, we set up VQE to solve a typical chemistry problem~\cite{34, 10, 2}~that has promising applications on an industry scale and is hard to solve classically: computing the ground state energy of a molecule.
Molecules consisting of H$_2$ or chains of four hydrogen atoms were investigated.
This experimental setup resembles~\cite{2}~where molecules consisting chains with up to 12 hydrogen atoms were computed on Google's Sycamore quantum computer.
H$_2$ emerged as a typical small-scale VQE benchmark and was analyzed in a number of quantum computing experiments~\cite{34, 10, 2}.

Each layer in the investigated RYRZ ansatz circuits (see fig.~\ref{9}) is constituted by parameterizable single qubit rotations and CNOT gates that entangle all qubits.
This structure was used in experiments on NISQ computers that showed the feasibility of VQE for small-scale problem instances~\cite{10}.
Two- and six-qubit RYRZ ansatz circuits were generated for computing the ground state energies of H$_2$ and chains of four hydrogen atoms respectively.
We investigated an increased number of layers in these ansatz circuits in our experiments to produce data about the scalability of VQE using larger quantum computations.
For larger-scale problems, VQE requires larger ansatz circuits with a larger number of parameters as the number of qubits increases~\cite{35, 13, 17, 15, 16, 14}.

We consider Pauli errors after each operation of a quantum gate on the quantum state as error model, whereas in~\cite{32}~errors before and after abstract unitary parameterizable layers are considered.
In the related work, noise is modeled using Pauli errors~\cite{21, 25}~or by modeling noise processes in a specific NISQ computer~\cite{28, 10}.
We chose the Pauli error model in this work since it was validated to be sufficiently accurate~\cite{9}~while not being too device-dependent:
device-specific noise processes such as crosstalk may not be indicative for different qubit technologies or different generations of quantum computers~\cite{3}.

\section{Results}\label{7}
In the following section, we first show the impact of Pauli errors on the solution obtainable by VQE and then focus on the error characteristics of the investigated ansatz circuits.
The last set of experiments shows convergence issues of VQE on medium-sized quantum computations subject to errors.

\subsection{Impact of Pauli Errors on VQE Solution}\label{8}
\begin{figure}[tb!]
  \centering

    \includegraphics[width=\linewidth]{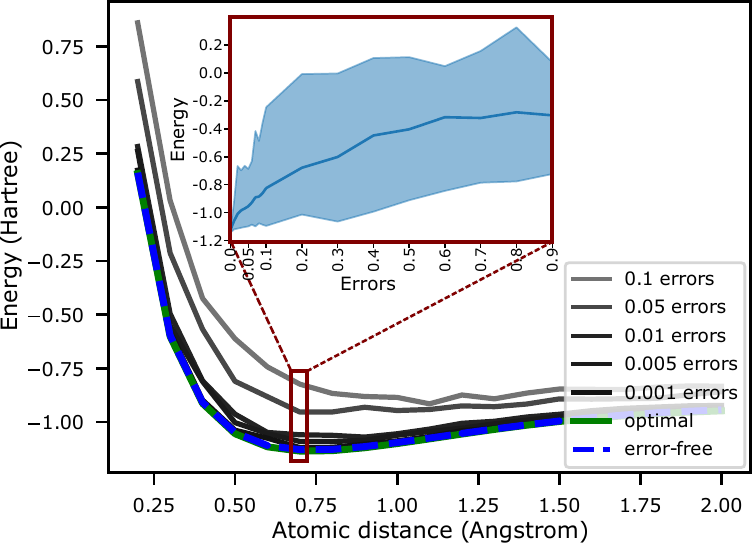}
  \caption{Ground state energy of H$_2$ at atomic distance $d=0.2,0.3,..,2.0$ angstrom as computed by an optimal solver and simulations with $0\%,...,0.1\%$ erroneous 2-qubit depth-7 RYRZ ansatz circuit computations.
  The inset shows average, and 5\% lower / upper bound of ground state energy for $d=0.7$ angstrom over 100 VQE runs when 0\% -- 90\% ansatz circuits of one VQE run are error-affected.\label{0}}
  \vspace{-2ex}
\end{figure}

In these experiments, the impact of errors on the overall solution computed by VQE was investigated.
VQE was used to compute the atomic distance at which the ground state energy of H$_2$ is minimal, i.e. the bond length.
All data points are averaged over 100 VQE computations and for atomic distance $d=0.7$ angstrom the average ground state energy and the 5\% upper and lower bound is reported in the inset of figure~\ref{0}.

Fig.~\ref{0}~shows the generated dissociation profiles subject to random Pauli errors.
For reference, the correct ground state energies are computed by a classical solver (optimal) and VQE with error-free ansatz circuit computations (error-free).
At most 0.1\% of ansatz circuits may incur one random Pauli error during their computation without impairing the ability of VQE to find the correct ground state energy at the correct bond length.
At 0.5\% and 1\% of erroneous ansatz circuits the ground state energy begins to deviate largely at 0.7 angstrom.
For 10\% erroneous ansatz circuits, the lowest ground state energy is found at an atomic distance of 1.1 angstrom.

At the atomic distance of 0.7 angstrom, the inset shows the average, and 5\% lower bound and upper bound of the 100 performed VQE runs with a varying ratio of erroneous ansatz circuits.
An increased variance of the computed ground state energy can be observed for an increasing ratio of erroneous ansatz circuits.
For instance, at 5\% erroneous ansatz circuits, the computed ground state energy varies from $-1.1$ to $-0.7$ Hartree, while at 0.1\% erroneous ansatz circuits the variance of the computed ground energy is insignificant.
Thus, computing a ground state energy with high confidence quickly requires a large number of measurements and repeated VQE runs with an increasing ratio of erroneous ansatz circuit computations.

Performing small-scale VQE, especially for computing the ground state energy of H$_2$, was demonstrated with high accuracy on contemporary NISQ computers~\cite{10, 12, 34}.
We stress that our results are in agreement and not in contradiction to published reports of physical experiments in~\cite{10, 12, 34}.
In fact, the graph in figure~\ref{0}~is almost identical to figure 3a of~\cite{10}.
For small problem instances, both physical measurements in~\cite{10}~and our predictions based on simulations are consistently successful.
However, our analysis extends to larger problem instances that are not covered by physical experiments conducted so far.
The good agreement on small problem sizes gives us confidence that our results have predictive power for larger-scale computations.
The findings in the next sections indicate a drop in accuracy if VQE is performed at the same physical error rate with larger ansatz circuits.

\subsection{Impact of Pauli Errors on RYRZ Ansatz Circuits}
\begin{figure}[tb]
  \centering

    \includegraphics[width=\linewidth]{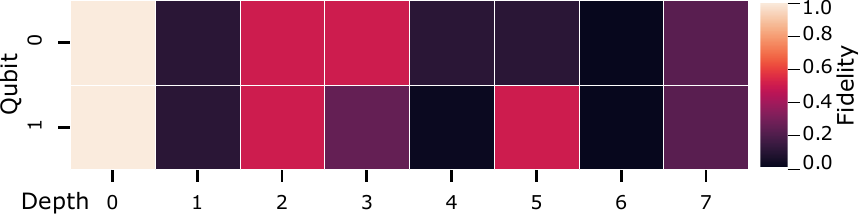}
  \caption{Impact of Pauli Z errors at different locations within the depth-8 2-qubit ansatz circuit of fig.~\ref{9}~with random parameters on fidelity. \label{2}}
  \vspace{-4ex}
\end{figure}

To examine the effect of errors on the investigated RYRZ ansatz circuits, an exhaustive Pauli error simulation was performed, i.e. each Pauli $X, Z, Y$ error at each possible error location in the ansatz circuit was simulated and the resulting fidelity reported.
Fig.~\ref{2}~reports on the fidelities of each Pauli Z error in the ansatz circuit of fig.~\ref{9}.
The x-axis shows on which quantum circuit depth and the y-axis on which qubit a Pauli Z error occurs.
While Pauli Z errors have no effect on the ansatz circuit fidelity before any gates are applied (i.e. depth zero), most of the remaining error locations lead to a fidelity close to zero.

The characteristics of the evaluated ansatz circuits are specified in table~\ref{12}.
H$_2$ was mapped to two qubits and the computation of five measurement circuits, while a chain of four hydrogen atoms was mapped to six qubits and 165 measurement circuits~\cite{27}.
The average fidelity of each error location and each Pauli error over 100 random parameters is $0.28$ and $0.279$ for the two and six qubit ansatz circuit with one layer respectively.
This drops to $0.073$ for the largest investigated 6-qubit ansatz circuit that was also investigated in section~\ref{3}.

Given the severity of an average Pauli error on ansatz circuits, the success of VQE subject to errors depends on the frequency of errors incurred during computations on a specific quantum computer and the impact of these errors on the classical processing, i.e. corruption of the optimization trajectory.

One random Pauli error per circuit computation happens on average, if the Pauli error rate of the computing NISQ device is $~9\%$ $(\approx 2.6\%)$ for the one-layer two- (six-) qubit circuits.
On the larger 20 layer circuits, the Pauli error rate required for one Pauli error on average is significantly reduced to $~0.7\%$ and $0.18\%$ for the two and six qubit circuit.
The lowest Pauli error rate for superconducting qubits at the time of writing is $0.15\%$ which implies that the largest ansatz circuit would be affected by at least one error on average during its computation~\cite{9}.

\subsection{Impact of Pauli Errors on VQE Convergence}\label{3}
\begin{figure*}[t]
\centering
\subfloat{\includegraphics[width=0.9\linewidth]{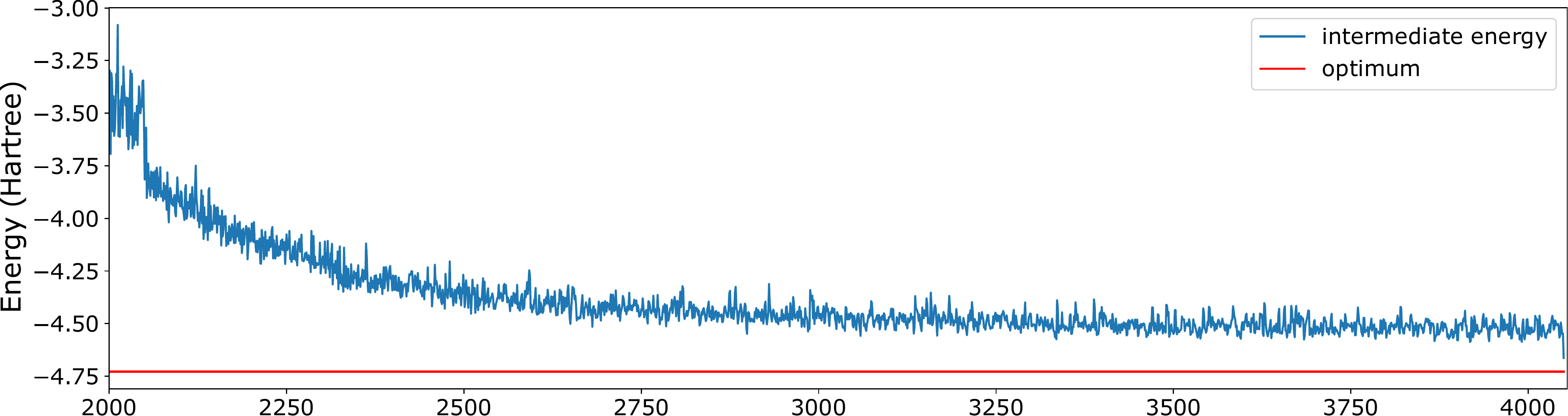}\label{a}}
\newline
\subfloat{\includegraphics[width=0.9\linewidth]{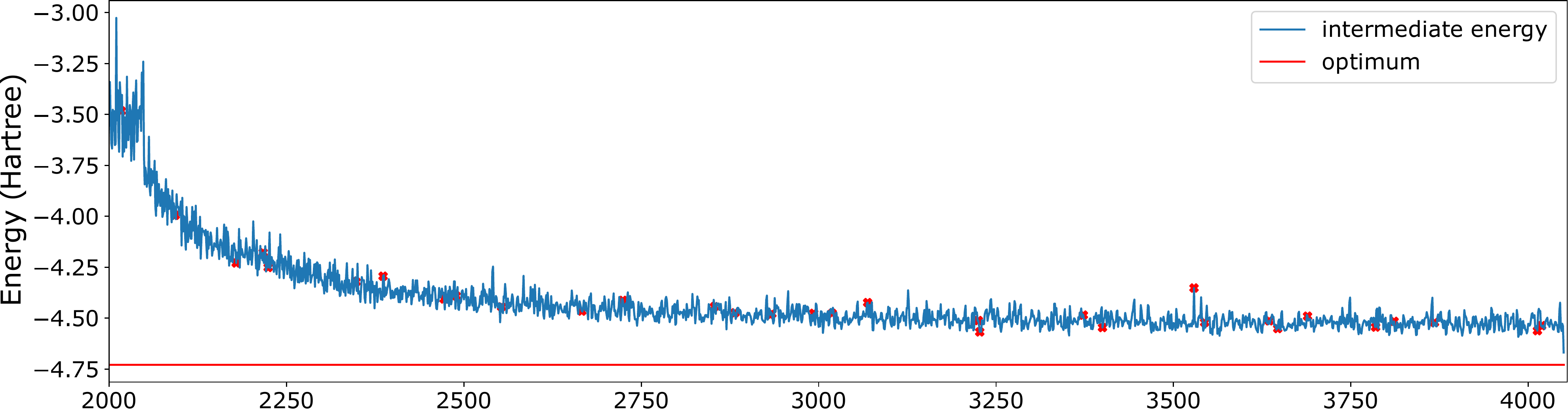}\label{b}}
\newline
\subfloat{\includegraphics[width=0.9\linewidth]{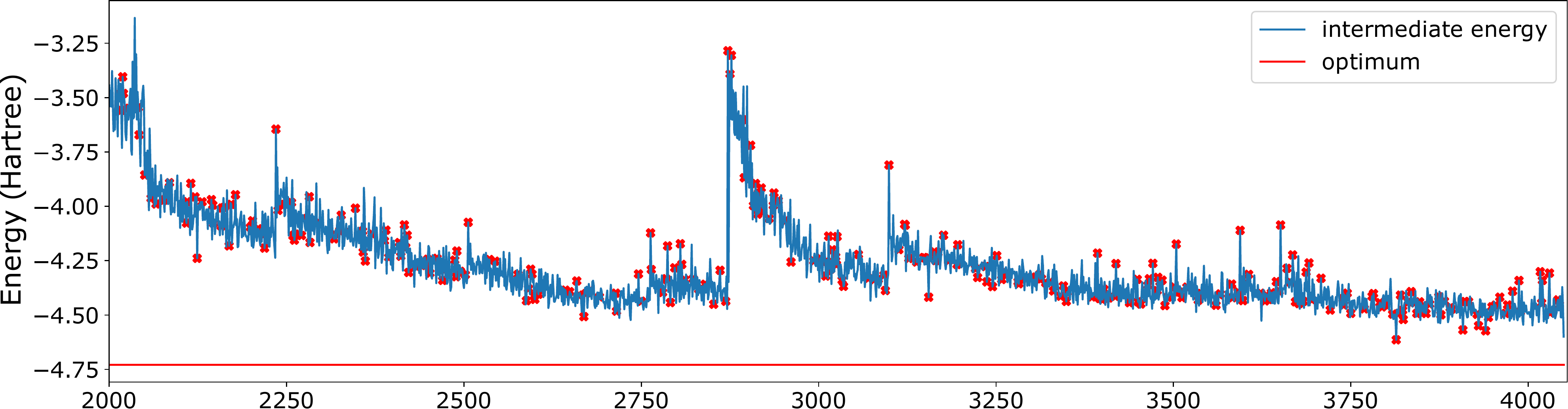}\label{c}}
\newline
\subfloat{\includegraphics[width=0.9\linewidth]{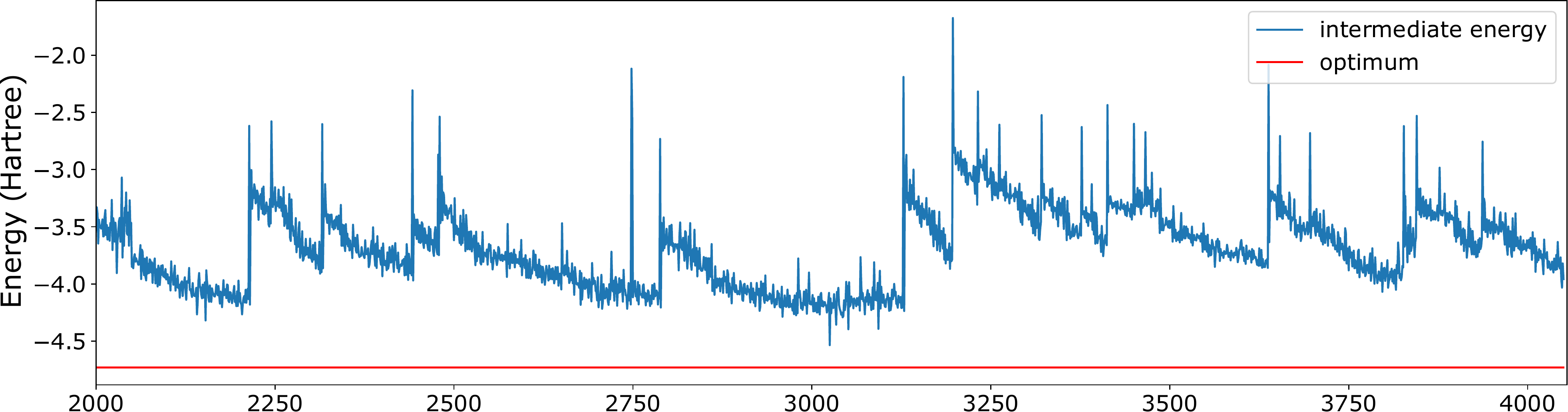}\label{d}}
\newline
\subfloat{\includegraphics[width=0.9\linewidth]{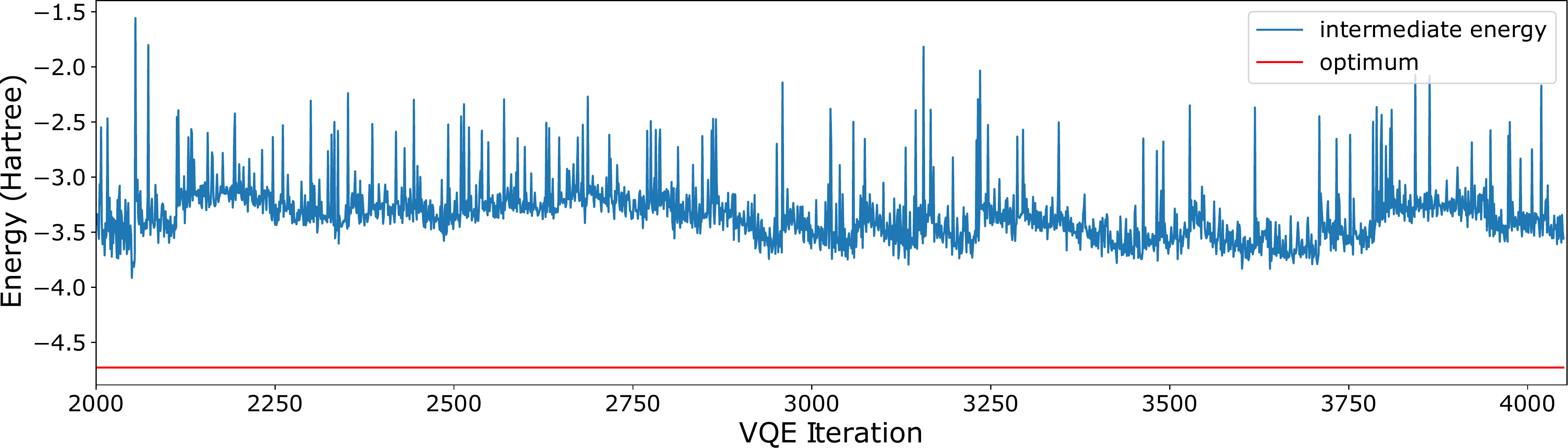}\label{e}}
\newline
\caption{Intermediate VQE solutions after 2000 error-free VQE iterations subject to $0\%, 0.01\%, 0.1\%, 1\%$ and $5\%$ erroneous 6-qubit 20-layer  RYRZ ansatz circuit computations (from top to bottom).
A red cross marks a VQE iteration with at least one erroneous ansatz circuit.
The optimum (ground state energy) is shown as a constant red line.\label{10}}
\end{figure*}

In this set of experiments the impact of Pauli errors on the convergence of VQE for a medium-sized ansatz circuit is highlighted in figure~\ref{10}.
VQE is first executed for roughly $2000$ iterations to provide an initial parameters for our experiments.
This enables the observation of VQE convergence close to the optimum and within its region of attraction in our experiments.

Then, VQE is simulated subject to Pauli errors starting on the initial parameters.
In particular, a fixed ratio of 6-qubit 20-layer RYRZ ansatz circuits during the VQE simulation are erroneous, i.e. are simulated subject to \emph{one} random Pauli error at a random location in the ansatz circuit.
For these VQE computations, the intermediate ground state energy solutions (y-axis) are displayed for each VQE iteration (x-axis)~in figure~\ref{10}.
The first graph in figure~\ref{10}~shows the convergence subject to no errors.
The second, third, fourth and fifth graphs are subject to $0.01\%,0.1\%, 1\%, 5\%$ erroneous ansatz circuits respectively.
For the graphs with $0.01\%$ and $0.1\%$ erroneous ansatz circuit computations, a red cross marks VQE iterations that contain erroneous computations.

While convergence close to the optimum is shown to be possible with $0\%$, $0.01\%$ and $0.1\%$ erroneous ansatz circuits in principle, convergence is completely restricted by $1\%$ or $5\%$ erroneous ansatz circuit computations.
The threshold of $0.1\%$ erroneous ansatz circuits convergence to the optimum is similar to the results on the 2-qubit depth-7 ansatz circuit in section~\ref{8} (figure~\ref{0}).
However, one Pauli error in every $100-th$ ansatz circuit computation requires that the computing NISQ device features a Pauli error rate that is significantly lower than specified in the last column of table~\ref{12}.
For the 6-qubit 20-layer ansatz circuit, the Pauli error rate of the computing NISQ device must be reduced by $98\%$ to yield the same ratio of erroneous ansatz circuits as with a 2-qubit depth-7 ansatz circuit.

As a consequence, active quantum error correction protocols, a significant reduction in the physical error rate of the computing device or a profound error suppression through quantum error mitigation methods will be a requirement for large-scale VQE applications.

\begin{table}[tb]
\footnotesize
\caption{Characteristics of evaluated RYRZ ansatz circuits. The last column contains the error rate that leads to one error per ansatz circuit on average\label{12}.}
\setlength{\tabcolsep}{1.3ex}

 \begin{tabular}{@{}rrrrrrrr@{}}
H & Layers & Qubits & Depth & \begin{tabular}[c]{@{}c@{}}Measure-\\ments\end{tabular} & \begin{tabular}[c]{@{}c@{}}Para-\\meters\end{tabular} & \begin{tabular}[c]{@{}c@{}}Average\\ Fidelity\end{tabular} & \begin{tabular}[c]{@{}c@{}}$ER$(\#F=1)\end{tabular} \\ \midrule
 2 & 1 & 2 & 7 & 5 & 8 & 0.280 & 0.0909 \\
 2 & 20 & 2 & 102 & 5 & 84 & 0.210 & 0.0069 \\
 4 & 1 & 6 & 15 & 165 & 24 & 0.279 & 0.0256 \\
 4 & 20 & 6 & 262 & 165 & 252 & 0.073 & 0.0018 \\ 
\end{tabular}
\vspace{-4ex}
\end{table}

\section{Conclusion: Consequences on NISQ}\label{6}
In this work, the Variational Quantum Eigensolver (VQE) was set up for a typical chemistry application~\cite{10}~and simulated subject to Pauli errors to examine the impact of errors on VQE.
The impact of errors is reported on three levels: fidelity of ansatz circuit computations, the overall VQE solution and the convergence of VQE.

We have shown that for basic ansatz circuits the fidelity drops significantly on an average Pauli error when the number of layers in the ansatz circuit is increased.
For a VQE computation requiring a six-qubit ansatz circuit with 20 layers, an error rate of less than $0.18\%$ is required and less than $1\%$ of ansatz circuit computations may be erroneous, otherwise errors deny convergence to the optimum.

Current error rates, in a variety of physical systems, are hovering around the 1\% rate.
Single qubit gates are generally of lower error, while qubit measurements errors remain slightly higher.
While it is not possible to predict how much lower physical qubit errors will be realized in the next 3-5 years, based on historical trends it would be surprising if average errors are suppressed lower than 0.01\% in any system~\cite{18}.

Theoretical studies have provided bounds for how noise-induced plateaus occur as VQE ansatz circuits are scaled to larger problem instances.
Our results demonstrate that the situation is even more dire than these theoretical models predict, especially if these variational algorithms are expected to be NISQ candidates. 
Even small problem instances with medium-sized ansatz circuits---those that do not need a quantum computer to solve---require error rates significantly lower than what is possible using current or near-term quantum computers.

These results dim expectations of successful computations using variational algorithms for molecules that require large ansatz circuits, i.e. molecules whose properties are intractable to compute classically.
Our results indicate for these computations that error correction protocols must be employed or the error rate of current quantum computers must drop significantly through error suppression or an improvement in quantum computing technology.

\bibliographystyle{IEEEtran}
\scriptsize{
\bibliography{vqe_errors}
}
\end{document}